\documentclass[11pt, a4paper]{article}
\usepackage[top=3cm, bottom=3cm, left=2.5cm, right=2.5cm]{geometry}
\usepackage[T1]{fontenc}
\usepackage{parskip}
\usepackage{amsmath}
\usepackage{amssymb}
\usepackage{amsthm}
\usepackage{mathtools}
\usepackage{bm}
\usepackage[pdfpagelabels,bookmarks=false]{hyperref}
\hypersetup{colorlinks, linkcolor=darkblue, citecolor=darkgreen, urlcolor=darkblue}
\usepackage[capitalize, nameinlink]{cleveref}
\usepackage{enumitem}
\usepackage{float}
\usepackage{caption}
\usepackage{subcaption}
\usepackage{tikz}
\usetikzlibrary{positioning}
\usetikzlibrary{calc}
\usetikzlibrary{decorations.markings}
\usetikzlibrary{arrows.meta}
\usepackage[linesnumbered,ruled,vlined]{algorithm2e}
\SetKwInput{KwInput}{Input}
\SetKwInput{KwOutput}{Output}
\DontPrintSemicolon
\foreach \x in {A,...,Z}{
\expandafter\xdef\csname b\x\endcsname{\noexpand\mathbb{\x}}
\expandafter\xdef\csname c\x\endcsname{\noexpand\mathcal{\x}}
\expandafter\xdef\csname f\x\endcsname{\noexpand\mathfrak{\x}}
}

\renewcommand{\hat}[1]{\widehat{#1}}

\newcommand{\abs}[1]{\left\lvert#1\right\rvert}

\newtheorem{theorem}{Theorem}
\newtheorem{lemma}[theorem]{Lemma}
\newtheorem{cor}[theorem]{Corollary}

\theoremstyle{remark}
\newtheorem{remark}[theorem]{Remark}
\theoremstyle{definition}
\newtheorem{definition}[theorem]{Definition}
\newtheorem{problem}[theorem]{Problem}

\usepackage[backend=bibtex,maxbibnames=6]{biblatex}
\usepackage{setspace}
\usepackage{colortbl}
\usepackage{tcolorbox}

\newcommand{\pccoalitionproblem}{\textsc{Prize-Collecting Coalition Problem}}
\newcommand{\pccoalitionnoproblem}{\textsc{Prize-Collecting Coalition}}
\newcommand{\subspacepccoalitionproblem}{\textsc{Subspace-Avoiding Prize-Collecting Coalition Problem}}
\newcommand{\subspacepccoalitionnoproblem}{\textsc{Subspace-Avoiding Prize-Collecting Coalition}}

\newcommand{\HN}{\mathcal{N}_h}
\newcommand{\Nuc}{\mathcal{N}}
\newcommand{\HNvalue}{\mathcal{V}_h}
\newcommand{\Nvalue}{\mathcal{V}}
\newcommand{\MPSvalue}{V}
\DeclareMathOperator{\excess}{excess}
\DeclareMathOperator{\opsize}{size}

\newcommand{\LC}{\mathrm{C}_l}
\newcommand{\HC}{\mathrm{C}_h}
\newcommand{\EHC}{\mathrm{C}^{\uparrow}_h}

\definecolor{darkblue}{rgb}{0,0,0.38}
\definecolor{darkred}{rgb}{0.8,0,0}
\definecolor{darkgreen}{rgb}{0.1,0.35,0}
\definecolor{lightblue}{rgb}{0.35,0.6,0.8}

\title{Nucleolus, Happy~Nucleolus, and Vehicle~Routing}
\author{Daniel Ebert \qquad Antonia Ellerbrock\\[.3cm] 
	\small \texttt{\{ebert,ellerbrock\}@dm.uni-bonn.de}\\[.3cm]
	\small Research Institute for Discrete Mathematics, University of Bonn}
\date{}

\begin{document}
\maketitle

\begin{abstract}
	We study the recently introduced fair division concept of the \emph{happy nucleolus} for cost allocation among players in a cooperative game,
	with special focus on its computation.
	The happy nucleolus applies the same fairness criterion as the well-established nucleolus but with reduced total value.
	Still, we show that the relation between the two concepts is quite involved, and intuitive properties do not hold --
	e.g., the entry of a player in the happy nucleolus can be larger than the entry of the same player in the nucleolus, even for monotone and subadditive games. This refutes conjectures of Meir et al.\ \cite{meir2011subsidies}.
	
	Further, we study the separation problem of the linear programs appearing in the MPS scheme for computing the (happy) nucleolus.
	It includes linear subspace avoidance constraints, which can be handled efficiently for problems with a certain
	dynamic programming formulation due to K\"ohnemann and Toth \cite{konemann2020general}. We show how to get rid
	of these constraints for \emph{all} monotone games if we allow for an arbitrarily small error of $\varepsilon$,
	thus conserving known approximation guarantees for the same problem without subspace avoidance.
	
	Finally, we focus on practical results at the example of vehicle routing games by designing an efficient heuristic based on our previous insights and past work, and demonstrate its power.
\end{abstract}

\section{Introduction}

We study \emph{cost allocations} for cooperative games. 

\begin{definition}[Cooperative game]
	A \emph{cooperative game} is a pair $(P,c)$ of a finite set $P$ of
	\emph{players} and a cost function $c: 2^P\to\bR$ on \emph{coalitions} $S \subseteq P$ with $c(\emptyset)=0$. $(P,c)$ is called
	\emph{monotone} if $c(S)\leq c(T)$ for all $S\subseteq T\subseteq P$, and \emph{subadditive} if $c(S\cup T)\leq c(S)+c(T)$ for all $S,T\subseteq P$ with $S\cap T=\emptyset$.
\end{definition}

Usually, $c$ is not given explicitly, but by some combinatorial optimization problem, e.g., as the size of a maximum matching in the subgraph $G[S]$ of some graph $G$ for $S \subseteq P$ \cite{deng1999algorithmic}.

\sloppy
Given a cooperative game $(P,c)$, a \emph{cost allocation} is a vector $y\in\bR^P$.
We will write \mbox{$y(S) \coloneqq \sum_{p \in S} y_p$} for $S \subseteq P$. 
There are multiple concepts defining 'fair' cost allocations, such as the \emph{Shapley value} \cite{shapley1953value}, the $\tau$-value \cite{tijs1981bounds} or the \emph{nucleolus} \cite{schmeidler1969nucleolus}. 
In this work, we will focus on the latter, introduced by Schmeidler in 1969 \cite{schmeidler1969nucleolus}, and a variant, the \emph{happy nucleolus}, introduced by Blauth et al.\ in 2024 \cite{blauth2024cost}.
For a formal definiton, let $\excess(S,y)\coloneqq c(S)-y(S)$ for $S \subseteq P$, $y\in\bR^P$.
The \emph{excess vector} $\theta(y)$ contains all values
$(\excess(S,y))_{S\subseteq P}$ in non-decreasing order.

\begin{definition}[Nucleolus \cite{schmeidler1969nucleolus}]
	\label{def:nuc}
	For a cooperative game $(P,c)$, the \emph{nucleolus} $\Nuc$ is the unique cost allocation
	that lexicographically maximizes the excess vector $\theta(y)$
	among all $y\in \bR^P$ with $y(P)=\Nvalue\coloneqq c(P)$.
\end{definition}

Not all coalitions need to be happy with the nucleolus:
$y \in\bR^P$ satisfies \emph{happiness}, also referred to as \emph{group rationality}, iff every coalition is \emph{happy} with it,
i.e., $y(S) \leq c(S)$ for all $S \subseteq P$.

\begin{definition}[Happy nucleolus \cite{blauth2024cost}]
	\label{def:happy_nuc}
	For a cooperative game $(P,c)$, the \emph{happy nucleolus} $\HN$ is the unique cost allocation
	that lexicographically maximizes the excess vector $\theta(y)$
	among all $y \in \bR^P$ with
	\[y(P) \ = \ \HNvalue \ \coloneqq \ \max\left\{ z(P) : z \in \bR^P \text{ satisfies happiness} \right\} \enspace .\]
\end{definition}

Nucleolus and happy nucleolus coincide iff $\Nvalue = \HNvalue$, or equivalently,
iff the \emph{core} of the game is non-empty:
The core $C$ consists of all cost allocations with total value $\Nvalue$ that satisfy happiness.

\subsection{Related Work and Our Results}

In this paper, we are interested in (efficient) computation of the happy nucleolus,
which we believe makes most sense when applied to games that incentivize cooperation. 
Thus, our focus will be on monotone and subadditive games.

For certain classes of games, the nucleolus can be computed in (pseudo-)polynomial time \cite{deng1994complexity,caprara2010new,chen2012computing,sziklai2017on,konemann2020computing,konemann2020general,pashkovich2022computing}, whereas for others, it is NP-hard \cite{faigle1998note,greco2015complexity}.
For so-called \emph{balanced} games with non-empty core (e.g.\ \cite{granot1981minimum,owen1975core,deng2009finding}), happy nucleolus and nucleolus coincide, and the results transfer.
Otherwise, NP-hardness of computing the happy nucleolus might be implied by NP-hardness of checking core emptiness or computing the so-called \emph{cost of stability} $\Nvalue-\HNvalue$ \cite{deng1998combinatorial,caprara2010new,liu2009complexity,chalkiadakis2016characteristic,biro2019generalized,bachrach2009cost,resnick2009cost,aziz2010monotone, chalkiadakis2016characteristic}. 
For \emph{set covering games}, Blauth et al.\ \cite{blauth2024cost} gave a polynomial-time algorithm, while computing the nucleolus remains NP-hard,
thus proving that the complexities of the two concepts can actually differ.

In \cref{sec:relation-nuc-happy-nuc}, we will discuss that the relation between nucleolus and happy nucleolus is quite involved.
We will give negative answers to conjectures by Meir et al.\ \cite{meir2011subsidies}.
Informally speaking and simplified, we will find that
\begin{itemize}
	\item the happy nucleolus is not dominated player-wise by the nucleolus,
	\item the happy nucleolus does not equal the nucleolus of the corresponding fractional game.
\end{itemize}

Moving on to computational aspects, our focus lies on the \emph{MPS scheme} that was found by Kopelowitz, Maschler, Peleg and Shapley for nucleolus computation early on \cite{kopelowitz1967computation,maschler1979geometric}, and later applied in various ways \cite{nguyen2016finding,engevall2004heterogeneous}.
It does not run in polynomial time in general, but needs to solve a linear number of linear programs with exponentially many constraints.
In \cref{sec:separation}, we will apply the MPS scheme to the happy nucleolus, and discuss the separation problem over its linear programs in detail, thereby showing how to get rid of certain linear subspace avoidance constraints for all monotone games at the expense of an arbitrarily small additive error.
Similarly, K\"onemann and Toth showed how to handle these constraints for all games that allow for a polynomial-sized dynamic program that solves the corresponding \pccoalitionproblem{} \cite{konemann2020general}.
Further, they proved that linear subspace avoidance constraints increase the complexity of some optimization problems from polynomial time solvability to NP-hardness.
However, for the \pccoalitionproblem{}, it is still open whether subspace avoidance can make the problem harder
(i.e., whether our additive error is necessary).

The gained insights of \cref{sec:separation} will be one building block of an efficient heuristic for computing the happy nucleolus  in \cref{sec:veh-routing-heuristic} on \emph{vehicle routing games}. 
Nucleolus computation has been studied on practical vehicle routing instances by  \cite{caprara2010new,gothe1996nucleolus,engevall2004heterogeneous}, to name only a few.
Guajardo and R\"onnqvist \cite{guajardo2016review}, as well as Schulte et al.\ \cite{schulte2019scalable} wrote surveys on cost allocation for these games.
We are not aware of any working heuristics for computing the (happy) nucleolus on large instances with hundreds of shipments -- a size needed for practical application -- and will close this gap.
We will provide practical results demonstrating the quality of our heuristic by comparing to the exact happy nucleolus on random instances with 50 players, and by showing convergence on large random instances, where we are not aware of an efficient method for computing the exact happy nucleolus to compare against.

We believe that our work naturally implies future research, see \cref{sec:future_research}.

\section{Relation Between Nucleolus and Happy Nucleolus}
\label{sec:relation-nuc-happy-nuc}

Nucleolus and happy nucleolus are quite similar cost allocation concepts. 
They use the same measure of fairness, but differ in total value.
We would like to understand their relation, 
possibly for deriving computational results.
Surprisingly, even simple and intuitive properties do not hold,
and we use this section to present two of them in detail.

\subsection{Player-Wise Domination and the Extended Happy Core}

As the happy nucleolus guarantees non-negative excess values everywhere, in particular for the grand coalition $P$, its total value $\HNvalue$ is never larger than the total value $\Nvalue = c(P)$ of the nucleolus.
It was open whether this domination holds player-wise, i.e., whether $\HN(p) \leq \Nuc(p)$ for all $p \in P$.
Intuitively, one might expect that when charging a reduced price total and applying the same fairness criterion, every single player should benefit.
In \cite{meir2011subsidies}, Meir, Rosenschein and Malizia conjectured this and related statements. 
We will give negative answers, demonstrating that the relation between $\Nuc$ and $\HN$ is quite involved.

\begin{definition}
	For a cooperative game $(P,c)$, the \emph{least core} $\LC$ contains all cost allocations $y$ with $y(P)=\Nvalue$ that violate happiness by at most $\varepsilon$, i.e., \mbox{$y(S) \leq c(S) + \varepsilon$} for all $S \subseteq P$, and $\varepsilon$ is chosen as the smallest number such that there exists such a cost allocation \cite{maschler1979geometric}.
	The \emph{happy core} $\HC$ contains all cost allocations $y$ with $y(P)=\HNvalue$ that satisfy happiness.
	The \emph{extended happy core} $\EHC$ contains all cost allocations $y$ with $y(P)=\Nvalue$ that dominate a point in the happy core player-wise, i.e., $y'_p \leq y_p$ for all $p \in P$ and some $y' \in \HC$. 
\end{definition}

It is clear that the nucleolus is always part of the least core, and the happy nucleolus is always part of the happy core: $\Nuc \in \LC$ and $\HN \in \HC$.
Meir et al.\ conjectured the inclusion $\LC \subseteq \EHC$
(cf.\ Conjecture 1 in \cite{meir2011subsidies}), and $\Nuc(p) \geq \HN(p)$ for all $p\in P$ (cf.\ Section 5.1 in \cite{meir2011subsidies}).
We will discuss an instance where we do not even have $\Nuc \in \EHC$.
Note that the properties of these core sets, nucleolus and happy nucleolus are equivalent for cost and profit sharing games under the transformation described in the following remark:

\begin{remark}[Value functions]
	Cooperative games can alternatively be defined over \emph{value functions} $\nu: 2^P\to\bR$ that are interpreted as gains, i.e., $\excess(S,y) = y(S) - \nu(S)$ for a coalition $S$ and a cost allocation $y$.
	We can transform a cost function $c$ into a value function $\nu \coloneqq - c_s$ with $c_s \coloneqq c(S) + \sum_{p \in P} s_p$ and player shifts $s_p \in \bR$ that are small enough such that $\nu \geq 0$.
	Then, $\excess_c(S,y) = \excess_{c_s}(S, y+s)$, and the (happy) nucleolus of $(P, \nu)$ results from the (happy) nucleolus of $(P,c)$ by adding the player shifts $s$, and vice versa.
\end{remark}

\begin{figure}[ht]
	\begin{center}
		\scalebox{.75}{
			\begin{tikzpicture}[xscale=1.5, yscale=1.3, ultra thick]
				\node[circle, fill, draw, scale = 0.5, label={[xshift=.05cm, yshift=-.65cm]{\scriptsize $p_1$}}] (1) at (1,0) {}; 
				\node[circle, fill, draw, scale = 0.5, label={[xshift=.05cm, yshift=-.65cm]{\scriptsize $p_2$}}] (2) at (3,0) {};
				\node[circle, fill, draw, scale = 0.5, label={[yshift=-.05cm]{\scriptsize $p_3$}}] (3) at (2,1.73) {};            
				
				\draw[darkblue] (1) -- (2) node[midway, fill=white] {\bm{$1$}};
				\draw[darkblue] (1) -- (3) node[midway, fill=white] {\bm{$1$}};
				\draw[darkblue] (2) -- (3) node[midway, fill=white] {\bm{$1$}};
				
				\begin{scope}[xshift=3cm]
					\node[circle, fill, draw, scale = 0.5, label={[xshift=.05cm, yshift=-.65cm]{\scriptsize $p_4$}}] (4) at (1,0) {}; 
					\node[circle, fill, draw, scale = 0.5, label={[xshift=.05cm, yshift=-.65cm]{\scriptsize $p_5$}}] (5) at (3,0) {};
					\node[circle, fill, draw, scale = 0.5, label={[yshift=-.05cm]{\scriptsize $p_6$}}] (6) at (2,1.73) {};        
					
					\draw[darkred] (4) -- (5) node[midway, fill=white] {\bm{$1$}};
					\draw[darkred] (4) -- (6) node[midway, fill=white] {\bm{$1$}};
					\draw[darkred] (5) -- (6) node[midway, fill=white] {\bm{$1$}};    
				\end{scope}
				
				\begin{scope}[xshift=6cm]
					\node[circle, fill, draw, scale = 0.5, label={[xshift=.05cm, yshift=-.65cm]{\scriptsize $p_7$}}] (7) at (1,0) {}; 
					\node[circle, fill, draw, scale = 0.5, label={[xshift=.05cm, yshift=-.65cm]{\scriptsize $p_8$}}] (8) at (3,0) {};
					\node[circle, fill, draw, scale = 0.5, label={[yshift=-.05cm]{\scriptsize $p_9$}}] (9) at (2,1.73) {};            
					
					\draw[darkblue] (7) -- (8) node[midway, fill=white] {\bm{$1$}};
					\draw[darkblue] (7) -- (9) node[midway, fill=white] {\bm{$1$}};
					\draw[darkblue] (8) -- (9) node[midway, fill=white] {\bm{$1$}};
				\end{scope}
				
				\draw[color=darkgreen] (3.5,.6) ellipse (90pt and 50pt) node[fill=white, yshift=65pt] {\bm{$3$}};
				\draw[color=darkgreen] (6.5,.6) ellipse (90pt and 50pt) node[fill=white, yshift=65pt] {\bm{$3$}};
			\end{tikzpicture}
		}
	\end{center}
	\caption{Set covering instance proving \cref{thm:counterexp_domination}. 
		Edges illustrate sets of size 2.\\
		We have $\HN \equiv \frac{1}{2}$, $\Nuc(p_i)=\frac{7}{15}$ for $i \in \{4,5,6\}$, and $\Nuc(p_i)=\frac{3}{5}$ otherwise.\\
		Especially, $\HN(p_4) > \Nuc(p_4)$. 
	}
	\label{fig:three_triangles}
\end{figure}
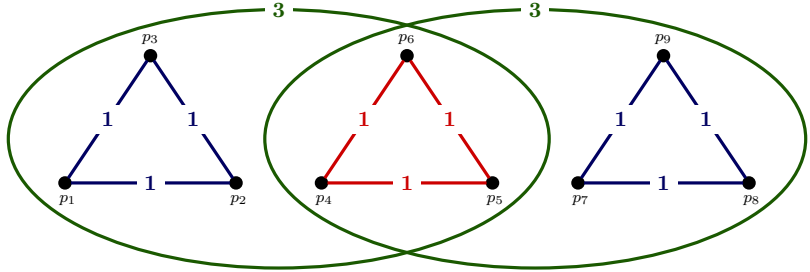

Recall that for set covering games, the cost of a coalition is determined by its minimum (not necessarily exact) covering cost.
In the classic triangle instance with unit cost, happy nucleolus and nucleolus differ: $\HN \equiv \frac{1}{2}$ and $\Nuc \equiv \frac{2}{3}$.
\cref{fig:three_triangles} uses three copies of this triangle, and proves \cref{thm:counterexp_domination}:

\begin{theorem}[Negative answers to conjectures in \cite{meir2011subsidies}] \label{thm:counterexp_domination}
	There is a monotone and subadditive game $(P,c)$, where $\Nuc \notin \EHC$, so in particular
	$\LC \not\subseteq \EHC$ and $\HN(p) > \Nuc(p)$ for some player $p \in P$.
\end{theorem}

\begin{proof}
	Consider the set covering instance depicted in \cref{fig:three_triangles}, where all input sets are drawn with their costs. 
	We have $\HN \equiv \frac{1}{2}$, and this is the unique solution that satisfies happiness with a total value of $\HNvalue = \frac{9}{2}$.
	On the other hand, the nucleolus attains a total value of $c(P)=5$, and balances the excess of the coalitions $P \setminus \{p_1\}, \{p_1,p_2,p_7,p_8\}$ and symmetric ones at $-\frac{2}{5}$.
	This determines $\Nuc(p_i)=\frac{3}{5}$ for $i \in \{ 1,2,3,7,8,9 \}$ and, by symmetry, fixes the remaining values to $\Nuc (p_i) = \frac{7}{15}$ for $i \in \{ 4,5,6 \}$.
	In particular, $\Nuc (p_4) = \frac{7}{15} < \frac{1}{2} = \HN (p_4)$.
	Further, $\HC = \left\{ \HN \right\}$.
	By $\Nuc (p_4) < \HN (p_4)$, $\Nuc \not \in \EHC$.
	With $\Nuc \in \LC$, this contradicts $\LC \subseteq \EHC$.
\end{proof}

\subsection{Fractional Costs}

We will see that the happy nucleolus of a game is not always equal to the nucleolus of the
corresponding fractional game, again by using set covering games.

\begin{definition}
	For a subadditive cooperative game $(P,c)$, its \emph{fractional game} $(P,c_f)$ is defined by \emph{fractional costs} $c_f$ for all coalitions $S \subseteq P$:
	\begin{equation*}
		\tiny
		c_f(S) \ := \ \min \left\{ \sum_{T \subseteq P} \alpha_T c(T) \ : \ \alpha_T \geq 0 \ \ \forall T \subseteq P \ , \ \sum_{p \in T \subseteq P} \alpha_T = 1 \ \ \forall p \in S \right\} \enspace .
	\end{equation*}
\end{definition}

\begin{figure}[ht]
	\begin{center}
		\scalebox{.7}{
			\begin{tikzpicture}[xscale=1.5, yscale=1.3, ultra thick]
				\node[circle, fill, draw, scale = 0.5, label={[xshift=.1cm, yshift=-.05cm]{\scriptsize $p_1$}}] (1) at (1.5,0) {}; 
				\node[circle, fill, draw, scale = 0.5, label={[xshift=-.35cm, yshift=-.3cm]{\scriptsize $p_2$}}] (2) at (0.3,-1) {};
				\node[circle, fill, draw, scale = 0.5, label={[xshift=-.35cm, yshift=-.3cm]{\scriptsize $p_3$}}] (3) at (0.3,1) {};       
				
				\begin{scope}[xshift=5cm]
					\node[circle, fill, draw, scale = 0.5, label={[xshift=-.1cm, yshift=-.05cm]{\scriptsize $p_4$}}] (4) at (-1.5,0) {}; 
					\node[circle, fill, draw, scale = 0.5, label={[xshift=.35cm, yshift=-.3cm]{\scriptsize $p_5$}}] (5) at (-.3,-1) {};
					\node[circle, fill, draw, scale = 0.5, label={[xshift=.35cm, yshift=-.3cm]{\scriptsize $p_6$}}] (6) at (-.3,1) {};            
				\end{scope}
				
				\draw[darkgreen] (3) -- (6) node[midway, fill=white] {\bm{$10$}}; 
				\draw[darkgreen] (1) -- (4) node[midway, fill=white] {\bm{$10$}}; 
				\draw[darkgreen] (2) -- (5) node[midway, fill=white] {\bm{$10$}};
				
				\draw[darkred] (1) -- (3) node[midway, fill=white] {\bm{$14$}};
				\draw[darkgreen] (1) -- (2) node[midway, fill=white] {\bm{$10$}};  
				\draw[darkgreen] (2) -- (3) node[midway, fill=white] {\bm{$10$}}; 
				
				\draw[darkred] (4) -- (6) node[midway, fill=white] {\bm{$14$}};
				\draw[darkgreen] (4) -- (5) node[midway, fill=white] {\bm{$10$}};  
				\draw[darkgreen] (5) -- (6) node[midway, fill=white] {\bm{$10$}}; 
				
				\draw[color=lightblue] (0.75,0) ellipse (40pt and 45pt) node[fill=white, yshift=58pt, xshift=-7pt] {\bm{$18$}}; 
			\end{tikzpicture}
		}
	\end{center}
	\caption{A set covering instance with a fractionally dominated set (blue), proving \cref{thm:frac_cost_function}.\\
		Edges illustrate sets of size~2. Sets with the same cost are drawn in the same color.
	}
	\label{fig:frac_dominated_sets}
\end{figure}
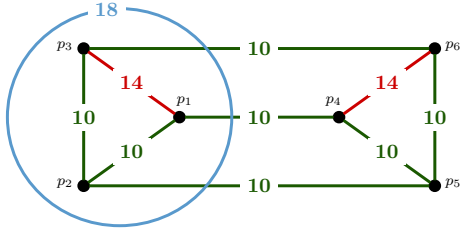

We call a set $S\subseteq P$ \emph{fractionally dominated} if $c_f(S) < c(S)$.
One might think that fractionally dominated sets are irrelevant for determining the happy nucleolus.
This would imply that the happy nucleolus of any game is equal to the nucleolus of the
corresponding fractional game since $\HNvalue(P,c)=c_f(P)$.
The example in \cref{fig:frac_dominated_sets} shows that this is not true in general.

\begin{theorem}\label{thm:frac_cost_function}
	There is a monotone and subadditive game $(P,c)$ such that 
	\[\HN(P,c) \ \neq \ \Nuc(P, c_f) \ = \ \HN(P, c_f) \enspace  .\]
	Additionally, there is a set $S\subseteq P$ with $c_f(S)<c(S)$ such that increasing or decreasing
	$c(S)$ changes $\HN(P,c)$.
\end{theorem}

\begin{proof}
	Consider the set covering instance depicted in \cref{fig:frac_dominated_sets}.
	Both $(P,c)$ and $(P,c_f)$ have a non-empty core, as shown by the cost allocation
	$y\equiv 5$ which satisfies happiness and has a total value of $30=c(P)=c_f(P)$.
	Additionally, any cost allocation that is in the core of one of the two games has
	excess 0 on all sets with cost 10 by LP duality: Each of these sets appears in some optimum
	fractional set covering of the instance. In particular, this is true for the
	(happy) nucleolus of both games.
	
	The (happy) nucleolus of $(P,c)$ then balances the excess values of the set with
	cost 18 and the set with cost 14 on the right. This results in
	\[
	\HN(P,c)_p \ = \ \Nuc(P,c)_p \ = \ \begin{cases}
		5+\frac{1}{3} & \text{ for } p \in \{p_2, p_4, p_6\} \enspace , \\
		5-\frac{1}{3} & \text{ for } p \in \{p_1, p_3, p_5\} \enspace ,
	\end{cases}
	\]
	and both sets having an excess value of $3+\frac{1}{3}$. It follows from the same calculation
	that increasing or decreasing the cost of the set with cost 18 changes the (happy) nucleolus of
	the instance.
	
	For the fractional cost function $c_f$, the instance is symmetric because the set
	with cost 18 is fractionally dominated. Therefore, the (happy) nucleolus is given by
	$\HN(P,c_f)=\Nuc(P,c_f)\equiv 5$. This is determined by balancing the excess values of
	the coalitions $\{p_1,p_2,p_3\}$ and $\{p_4,p_5,p_6\}$, both with fractional cost 17.
\end{proof}

\section{Separating over the Polytopes in the MPS Scheme}
\label{sec:separation}

In this section, we study the Maschler-Peleg-Shapley (MPS) scheme which can be used to compute the nucleolus and the happy nucleolus.
We show how the main subproblem that has to be solved can be reduced to an easier special case.
For technical reasons, we assume that $\max_{S\subseteq P}\opsize(c(S))$ is polynomially bounded,
where $\opsize(\cdot)$ denotes the bit complexity of the given number. This assumption is trivially true for
all interesting problems we are aware of.

\subsection{The MPS Scheme}
\label{subsec:mps}

As shown by Maschler, Peleg and Shapley \cite{maschler1979geometric}, the nucleolus can be computed by solving a sequence of linear programs. 
An improved variant which needs at most a linear number of iterations was proposed by \textcite{solymosi1993computing}.
It trivially extends to the happy nucleolus, see \cref{fig:mps}.

\begin{figure}[ht]
	\begin{tcolorbox}[
		colback=gray!10, 
		halign=left, 
		rounded corners, 
		colframe=gray!50, 
		boxrule=1.5pt 
		]
		\begin{description}
			\item[Step 1:] Let $\cS_\text{fixed} \gets \emptyset$ and $\MPSvalue \gets \Nvalue$ (for $\Nuc$) or $\MPSvalue \gets \HNvalue$ (for $\HN$).\\[.2cm]
			\item[Step 2:] Compute a pair of optimum primal and dual solutions $((\xi, y), z)$ to \eqref{eq:mps-scheme-lp}, where $\mathrm{span}(\cS_\text{fixed})$ denotes the linear subspace of $\bR^P$ generated by the incidence vectors of coalitions in $\cS_\text{fixed}$.
			\begin{equation}
				\small 
				\begin{alignedat}{3}
					&\max\mathrlap{\ \xi} \\
					&\ \text{s.t.}\quad& y(S)&\ = \ c(S) - \xi^*_S
					\qquad &&\forall S\in\cS_\text{fixed} \\
					&&y(S) + \xi &\ \leq \ c(S)
					\qquad &&\forall S\in 2^P\setminus
					\mathrm{span}(\cS_\text{fixed}) \\
					&&y(P)&\ = \ \MPSvalue\\
					&&\xi &\ \in \ \bR \\
					&&y &\ \in \ \bR^P
				\end{alignedat}\label{eq:mps-scheme-lp}
			\end{equation}
			\item[Step 3:] For each $S\in 2^P\setminus\mathrm{span}(\cS_\text{fixed})$ with $z_S \neq 0$, $\cS_\text{fixed} \gets \cS_\text{fixed} \cup \{S\}$, $\xi^*_S \gets \xi$.\\[.2cm]
			\item[Step 4:] Repeat Steps 2 and 3 until $\mathrm{span}(\cS_\text{fixed})=\bR^P$.
			Return $y$.
		\end{description}
	\end{tcolorbox}
	\caption{MPS Scheme}
	\label{fig:mps}
\end{figure}

\begin{theorem}[Maschler, Peleg and Shapley \cite{maschler1979geometric}, Solymosi \cite{solymosi1993computing}]
	The scheme described in \cref{fig:mps} correctly computes the (happy) nucleolus in at most $\abs{P}$ iterations. \qed
\end{theorem}

In order to apply this scheme, one needs to be able to compute the total value $\MPSvalue$ and solve the linear program (\ref{eq:mps-scheme-lp}).
All other steps are trivial. 
For the happy nucleolus (in contrast to the nucleolus), computing $\MPSvalue = \HNvalue$ can also be reduced to
solving (\ref{eq:mps-scheme-lp}) because $\HNvalue$ is the maximum value for which the optimum value of (\ref{eq:mps-scheme-lp}) with $\cS_\text{fixed} = \emptyset$ is nonnegative (and in fact $0$):
$\opsize(\HNvalue)$ is bounded polynomially\footnote{
	The bound on $\opsize(\HNvalue)$ comes from the fact that $\HNvalue$ is the optimum solution value of a linear program with $\abs{P}$
	variables and coefficients with bit complexity $\max_{S\subseteq P}\opsize(S)$.
}, so the exact value of $\HNvalue$ can be computed
by binary search \cite{papadimitriou1979efficient}.

If the constraints $y(S)+\xi\leq c(S)$ for $S\in 2^P\setminus \mathrm{span}(\cS_\text{fixed})$ can be separated efficiently, then
one can also solve (\ref{eq:mps-scheme-lp}) and its dual \cite{grotschel1981ellipsoid}, and therefore compute the happy nucleolus.
For this reason, we focus on the separation problem for this family of constraints in the remainder of this section.

\subsection{The Subspace-Avoiding Prize-Collecting Coalition Problem}

The separation problem over the non-trivial constraints of \eqref{eq:mps-scheme-lp} can be reformulated as follows for nonnegative cost allocations. Note that for monotone cost functions, both the nucleolus and the happy nucleolus are always nonnegative.

\begin{problem}[\subspacepccoalitionnoproblem{}]
	\label{problem:subspace-avoiding-prize-collecting-coalition}
	Given a cooperative game $(P,c)$, prizes $\pi\in\bR^P_{\geq0}$ and a linear
	subspace $L\subsetneq \bR^P$,
	find a coalition $S \in 2^P \setminus L$ and a cost estimate $\lambda\geq c(S)$ with $\lambda+\pi(P\setminus S)$ minimum.
\end{problem}

We will reduce the \subspacepccoalitionproblem{} to a simple prize-collecting problem which essentially corresponds to the special case $L=\{0\}$.
Our reduction will add an arbitrarily small additive error.

\begin{problem}[\pccoalitionnoproblem{}]
	\label{problem:prize-collecting-coalition}
	Given a cooperative game $(P,c)$ and prizes $\pi\in\bR^P_{\geq0}$, find a coalition
	$S\in 2^P$ and a cost estimate $\lambda\geq c(S)$
	with $\lambda + \pi(P\setminus S)$ minimum.
\end{problem}

\begin{lemma}\label{lem:price-collecting-coalition-restr}
	Let $\cF$ be a family of monotone cooperative games and $\alpha\geq1$
	such that there is an $\alpha$-approximation algorithm for the
	\pccoalitionproblem{} for $\cF$.
	Then, given $(P,c)\in\cF$, prizes $\pi\in\bR^P_{\geq0}$, and
	disjoint subsets $A,B\subseteq P$, we can find an $\alpha$-approximate solution to the
	\pccoalitionproblem{} restricted to coalitions $S$ with $A\subseteq S
	\subseteq P\setminus B$ in polynomial time.
\end{lemma}

\pagebreak

\begin{proof}
	Let $(P,c, \pi, A, B)$ be an instance of the restricted
	\pccoalitionproblem{} with $(P,c)\in\cF$.
	For $p \in P$, we define
	\[
	\hat \pi_p \ \coloneqq \ \begin{cases}
		\alpha\cdot U+1 \qquad&\text{ if }p\in A \enspace , \\
		0 \qquad&\text{ if }p\in B \enspace , \\
		\pi_p \qquad&\text{ otherwise} \enspace ,
	\end{cases}
	\]
	where $U\geq0$ is an upper bound\footnote{
		The existence of such a bound with polynomial size follows from the technical assumption that $\opsize(c(P))$ is bounded.
		Note that we do not need to know the bound on $\opsize(c(P))$ a-priori. Instead, we can simply start with an arbitrary value
		for $U$ and, if this was too small, double $U$ and restart the procedure.
	} on $c(P)$.
	Let $(S',\lambda')$ be
	an $\alpha$-approximate solution of the
	\pccoalitionproblem{} $(P,c,\hat \pi)$. 
	Then, $A\subseteq S'$ (otherwise $S'$ cannot be an $\alpha$-approximate
	solution by definition of $\hat \pi$), so by monotonicity of $c$,
	$(S'',\lambda')$ with $S''\coloneqq S'\setminus B$ is a feasible
	solution to the restricted problem.
	
	Let $S^*$ be an optimum solution to the restricted problem for $\pi$. Then,
	\begin{align*}
		\lambda' \ + \ \pi \left( P \setminus S'' \right) \
		&= \ \lambda' \ + \ \hat \pi \left( P\setminus S'' \right) \ + \ \pi(B) \\[.1cm]
		&\leq \ \alpha\cdot \big( c(S^*) \ + \ \hat \pi \left( P\setminus S^* \right) \big) \ + \ \pi(B) \\[.1cm]
		&\leq \ \alpha\cdot \big( c(S^*) \ + \ \hat \pi \left( P\setminus S^* \right) \ + \ \pi(B) \big) \\[.1cm]
		&= \ \alpha\cdot \big( c(S^*) \ + \ \pi \left( P\setminus S^* \right) \big) \enspace ,
	\end{align*}
	so $\left( S'',\lambda' \right)$ is an $\alpha$-approximate solution to the
	restricted problem for $\pi$.
\end{proof}

\begin{theorem}\label{thm:red-subspace-avoiding-coal}
	Let $\cF$ be a family of monotone cooperative games and $\alpha\geq1$
	such that there exists an $\alpha$-approximation algorithm for the
	\pccoalitionproblem{} for $\cF$.
	Then, there exists an
	$(\alpha+\varepsilon)$-approximation algorithm
	for the \subspacepccoalitionproblem{} for $\cF$
	for any fixed $\epsilon > 0 $.
\end{theorem}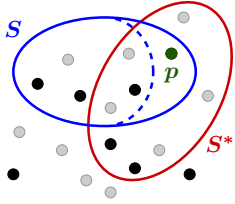
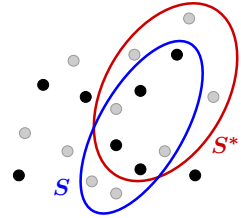
\begin{figure}[ht]
	\newcommand{\ExampleShared}{
		\begin{scope}[every node/.style={circle,fill,draw,scale=0.5}]
			\node (p1) at (0,0) {};
			\node (p2) at (1.1,1.3) {};
			\node (p3) at (2,0.1) {};
			\node (p4) at (2,1.4) {};
			\node (p5) at (2.6,2) {};
			\node (p6) at (0.4, 1.5) {};
			\node (p7) at (1.6, 0.5) {};
			\node (p8) at (2.9, 0) {};
		\end{scope}
		\begin{scope}[every node/.style={circle,fill=gray!40,draw=gray!80,scale=0.5}]
			\node (p'1) at (0.1,0.7) {};
			\node (p'2) at (0.8,0.4) {};
			\node (p'3) at (0.9,1.9) {};
			\node (p'4) at (1.6,1.1) {};
			\node (p'5) at (1.9,2) {};
			\node (p'6) at (2.4, 0.4) {};
			\node (p'7) at (1.6, -0.3) {};
			\node (p'8) at (3.2, 1.3) {};
			\node (p'9) at (2.8, 2.6) {};
			\node (p'10) at (1.2, -0.1) {};
		\end{scope}
		\draw[very thick,darkred,rotate=60] (2.4,-1.4) ellipse (1.6 and 1);
		\node[darkred] at (3.4,0.5) {\bm{$S^*$}};
		\node at (1,-0.4) {};
	}
	\begin{subfigure}[t]{0.45\textwidth}
		\begin{center}
			\scalebox{.8}{
				\begin{tikzpicture}
					\ExampleShared
					\node[circle,draw,fill,scale=0.5,color=darkgreen] (p) at (p5) {};
					\node[color=darkgreen] [below=0 of p] {\bm{$p$}};
					\draw[very thick,blue] (1.5,1.7) ellipse (1.5 and 0.9);
					\begin{scope}
						\clip (1.6,0) rectangle (3,3);
						\draw[very thick,blue,dashed] (1.5,1.7) ellipse (0.8 and 0.9);
					\end{scope}
					\node[blue] at (0,2.4) {\bm{$S$}};
			\end{tikzpicture} }
			\subcaption{
				\bm{$\textcolor{darkgreen}{p}\in \left( \textcolor{darkred}{S^*}\cap P' \right)$}
				satisfies (\ref{eq:subspace-avoid-pen-bound}). 
				\textcolor{blue}{\bm{$S$}}
				and \bm{$\textcolor{blue}{S} \setminus \{ \textcolor{darkgreen}{p} \}$} are cheap and at least one of them
				is feasible.
			}
		\end{center}
	\end{subfigure}
	\hfill
	\begin{subfigure}[t]{0.45\textwidth}
		\begin{center}
			\scalebox{.8}{
				\begin{tikzpicture}
					\ExampleShared
					\draw[very thick,blue,rotate=60] (1.7,-1.35) ellipse (1.6 and 0.7);
					\node[blue] at (0.7,-0.2) {\bm{$S$}};
			\end{tikzpicture} }
			\subcaption{
				No \bm{$p \in \left( \textcolor{darkred}{S^*}\cap P' \right)$} satisfies (\ref{eq:subspace-avoid-pen-bound}).
				\textcolor{blue}{\bm{$S$}} is cheap and feasible since it agrees with \bm{$\textcolor{darkred}{S^*}$} on \bm{$P'$}.
			}
		\end{center}
	\end{subfigure}
	\caption{
		An illustration of the proof of Theorem \ref{thm:red-subspace-avoiding-coal}.
		Players in $\bm{P'}$ are drawn as black points, all others gray.
		\textcolor{darkred}{$\bm{S^*}$} is an optimum solution to the subspace-avoiding problem.
		The solution(s) found by our algorithm are shown in blue.
	}
	\label{fig:red-subspace-avoiding-coal}
\end{figure}
\begin{proof}

	Let $(P,c, \pi, L)$ be an instance of the
	\textsc{Subspace-Avoiding Prize-Col-lecting Coalition Problem} with $(P,c)\in\cF$.
	Let $P'\coloneqq \left\{ p\in P : {\{p\}}\notin L\right\}$ and
	$K\coloneqq \big\lceil \frac{1}{\varepsilon} \big\rceil$.
	Note that $P'\neq\emptyset$.
	Let $(S^*, c(S^*))$ be an optimum solution. We proceed differently depending on whether there exists a player $p\in \left( S^*\cap P' \right)$ with low prize
	\begin{equation}
		\pi_p \ \leq \ \varepsilon\cdot\pi(P\setminus S^*) \enspace . 
		\label{eq:subspace-avoid-pen-bound}
	\end{equation}
	If not, we want to work with a certain 'out-set' $O \subseteq (P'\setminus S^*)$ of bounded size $|O| \leq K$.
	Even though we have to try all possibilities for $p$ and $O$ (because we do not know $S^*$), the procedure takes only polynomial time because $K$ is constant.
	\cref{fig:red-subspace-avoiding-coal} illustrates the following two cases.
	
	\begin{itemize}
		\item[(a)] $p \in \left( S^*\cap P' \right)$ satisfies (\ref{eq:subspace-avoid-pen-bound}).\\
		Let $(S,\lambda)$ be an $\alpha$-approximate solution to the \pccoalitionproblem{} restricted to coalitions containing $p$. This can be computed in polynomial time by Lemma \ref{lem:price-collecting-coalition-restr}. 
		By $\{p\}\notin L$, we have $S\notin L$ or $(S\setminus\{p\})\notin L$. Additionally, both solutions are sufficiently cheap, so we can simply return whichever is feasible:
		\begin{alignat*}{4}
			\pi(P\setminus S) \ + \ \lambda 
			&\ \leq \ & \alpha\quad \cdot\quad&\big( \pi(P\setminus S^*) \ + \ c(S^*) \big) \enspace , \\[.2cm]
			\pi(P\setminus (S\setminus\{p\})) \ + \ \lambda
			&\ = \mathrlap{\ \pi_p\ + \ \pi(P\setminus S) \ + \ \lambda} \\
			&\ \leq \ &\; (\alpha+\varepsilon)\;\cdot\;&\left( \pi(P\setminus S^*) \ + \ c(S^*) \right) \enspace ,
		\end{alignat*}
		where we used that $S^*$ is a feasible solution to the restricted problem, $\lambda$ remains valid for $S \setminus \{ p \}$ by monotonicity of $(P,c)$, and $p$ satisfies \eqref{eq:subspace-avoid-pen-bound}.
		\item[(b)] No $p \in \left( S^*\cap P' \right)$ satisfies (\ref{eq:subspace-avoid-pen-bound}).\\
		We assume to have found $O \subseteq (P'\setminus S^*)$ of size $|O| =  \min\{K, \abs{P'\setminus S^*}\}$ that maximizes the prize $\pi(O)$,
		and define the 'in-set'
		\[
		I\ \coloneqq \ 
		\begin{cases}
			\big\{ p\in \left( P'\setminus O \right) : \pi_p>\min_{q\in O}\pi_q \big\}
			\qquad &\text{ if }\abs{O}=K \enspace , \\[.2cm]
			P'\setminus O &\text{ if }\abs{O}<K \enspace .
		\end{cases}
		\]
		We claim that $S^*\cap P'=I$.
		This is trivial for $\abs{O} < K$ by definition of $O$ and $I$.
		Otherwise, $I \subseteq (S^* \cap P')$ because $O$ maximizes $\pi(O)$ inside $P'\setminus S^*$.
		Additionally, $(S^*\cap P') \subseteq I$, because for every $p \in (S^*\cap P')$, \eqref{eq:subspace-avoid-pen-bound} does not hold, and thus
		\[
		\pi_p \ > \ \varepsilon \cdot \pi \left( P\setminus S^* \right) \ 
		\geq \ \frac{1}{K} \cdot \pi \left( P'\setminus S^* \right) \ 
		\geq \ \frac{1}{K} \cdot \pi(O) \
		\geq \ \min_{q \in O} \pi_q \enspace .
		\] 
		Let $(S,\lambda)$ be an $\alpha$-approximate solution to the \pccoalitionproblem{} restricted to coalitions $S$ with $S\cap P'=S^*\cap P'=I$, again using Lemma \ref{lem:price-collecting-coalition-restr}.
		Then, $S^*\notin L$ implies $S\notin L$, so $S$ is feasible. 
		Additionally, $\pi(P\setminus S)+\lambda \leq \alpha\cdot(\pi(P\setminus S^*)+c(S^*))$ because $S^*$ is a feasible solution to the restricted problem. \qedhere
	\end{itemize}
\end{proof}

With \cref{thm:red-subspace-avoiding-coal}, we can apply known approximation algorithms for prize-collecting
problems to their subspace-avoiding variants, achieving essentially the same approximation ratios:

\begin{cor}[based on \cite{blauth2025better}]
	There is a 1.599-approximation algorithm for the \textsc{Symmetric Subspace-Avoiding Prize-Collecting TSP}. \qed
\end{cor}

\begin{cor}[based on \cite{ahmadi2024prize}]
	There is a 1.7995-approximation algorithm for the \textsc{Subspace-Avoiding Prize-Collecting Steiner Tree Problem}. \qed
\end{cor}

\section{Heuristic Computation of the Happy Nucleolus for Vehicle Routing Games}
\label{sec:veh-routing-heuristic}

Because of its practical relevance for postal services and certain other service providers, we will discuss heuristic computation of the happy nucleolus on vehicle routing games in this section,
combining our insights from \cref{sec:separation} with \cite{blauth2024cost} and \cite{packing-lp-solver} and further speed-ups.

\begin{definition}
	\sloppy
	\emph{Vehicle routing games} are defined over a complete graph \mbox{$G=\left( P \cup \{d\}, E , w \right)$} with nonnegative edge weights $w$
	and certain constraints defining coalitions $T \subseteq P$ that can be served together and will be called \emph{tours} in the following.
	For example, there could be a fixed maximum number of players per tour, or a weight constraint.
	The cost $c(S)$ of a coalition $S \subseteq P$ is defined as the minimum total edge weight of a set of walks in $G$, where each walk starts and ends at the \emph{depot} vertex $d$ and is otherwise restricted to some tour $T \subseteq P$, and each member of $S$ is visited by at least one walk.
\end{definition}

In terms of their cost function, vehicle routing games are equivalent to set covering games with all tours and their cost as input sets. 
Thus, the happy nucleolus of vehicle routing games is fully determined by the excess values of all tours by \cite{blauth2024cost}.
However, vehicle routing games can usually be encoded in size polynomial in $\abs{P}$, so the complexity results of \cite{blauth2024cost} cannot be transferred.
Of course, for \emph{fixed capacity} vehicle routing games, where $T \subseteq P$ is a tour if and only if $\abs{T} \leq k$ for some given constant $k$, the happy nucleolus can be computed in polynomial time as the number of tours is polynomial in $\abs{P}$.
But even in this easy case, exact computation would be too slow for practical usage.

In the following, we will work with general vehicle routing games, where happy nucleolus computation is NP-hard by reduction from \textsc{TSP}.
\cref{subsec:heuristic} describes our heuristic, and \cref{subsec:results} demonstrates practical results on randomly generated instances.

\subsection{Our Heuristic}
\label{subsec:heuristic}

In order to heuristically compute the happy nucleolus in practice, we use a constraint generation approach.
We maintain a cost allocation $y\in\bR^P$ and a set $\cT \subseteq 2^P$ of coalitions
that might be relevant for determining the happy nucleolus, motivated by the fact that
there always exists a set of at most $2(\abs{P}-1)$ coalitions that fully determines the happy nucleolus\footnote{
	For vehicle routing games, such a set is of course NP-hard to identify.
} \cite{reijnierse1998b}.
Our heuristic is outlined in \cref{fig:heuristic}, and details for steps 2) and 3) are given below.

\begin{figure}[ht]
	\begin{tcolorbox}[
		colback=gray!10, 
		halign=left, 
		rounded corners, 
		colframe=gray!50, 
		boxrule=1.5pt 
		]
		\begin{description}
			\item[1) Initialization:] Start with \ $\cT \gets \emptyset$ , \ $\cL \gets \emptyset$ \ and \ $y_p \gets$ c(\{p\}).\\[.2cm]
			\item[2) Tour generation:] Add a diverse set of coalitions to $\mathcal{T}$ that have small excess with respect to $y$, and avoid the linear subspaces in $\cL$  (by heuristically separating over \eqref{eq:mps-scheme-lp}).\\[.2cm]
			\item[3) Computing a cost allocation:] Run the MPS scheme with a modified LP and restricted to $\cT$, save the result in $y$,
			and let $\cL$ be the set of linear subspaces $\mathrm{span}(\cS_\text{fixed})$ of all iterations.\\[.2cm]
			\item[4) Repeat:] Alternate steps 2) and 3). Occasionally, prune $\cT$ by deleting coalitions that have not been relevant for some time.
			Stop after a fixed number of repetitions.
		\end{description}
	\end{tcolorbox}
	\caption{Sketched happy nucleolus heuristic for vehicle routing games}
	\label{fig:heuristic}
\end{figure}

\subsubsection{Tour Generation:} 

Given a cost allocation $y$, we want to find tours $T$ that minimize the excess $c(T)-y(T)$
while avoiding the linear subspaces $L \in \cL$ that appeared when computing $y$ with a variant of the MPS scheme.
We cannot hope to do this optimally for general $L$ on larger instances, 
as this would imply identification of violated constraints of \eqref{eq:mps-scheme-lp} (if there are any),
which in general is at least as hard as prize-collecting TSP.
Instead, we use a simple heuristic. 

Our procedure is very similar to Case (a) of the proof of
Theorem \ref{thm:red-subspace-avoiding-coal}.
The only differences are that we greedily compute minimum-excess tours instead of using
an $\alpha$-approximation,
and that we consider player insertions next to removals. 
We start by computing a set of tours that covers all players, greedily minimizing
the excess of each tour, and add this set to $\cT$. Then, we consider the set of
tours resulting from removing/inserting a single player from/into a tour in $\cT$.
For each subspace $L\in\cL$, we select a minimum-excess tour that avoids $L$
from this candidate set, and add it to $\cT$.

\subsubsection{Computing a Cost Allocation:} 

Given a set of tours $\cT$ and a total value $V$, a cost allocation that lexicographically maximizes the excess on $\cT$ can be
computed in polynomial time by the MPS scheme. However, this includes solving up to $\abs{P}$
linear programs which is challenging to do in reasonable time on large instances, 
even if $\cT$ is comparatively small (in our case, $\abs{\cT} \in \mathcal{O}(\abs{P})$).
For this reason, we modify the linear program \eqref{eq:mps-scheme-lp} to \eqref{eq:mps-scheme-packing-lp} with sufficiently large $\gamma$ (where $\gamma=2$ already produces good results in practice):
\begin{equation}
	\small 
	\begin{alignedat}{3}
		&\max\mathrlap{\ \gamma\cdot y(P) + \xi} \\[.1cm]
		&\ \text{s.t.}\quad& y(S)&\ \leq \ c(S) - \xi^*_S
		\qquad &&\forall S\in\cS_\text{fixed} \\
		&&y(S) + \xi &\ \leq \ c(S)
		\qquad &&\forall S\in \cT\setminus \cS_\text{fixed} \\
		&&\xi &\ \geq \ 0 \\
		&&y &\ \in \ \bR^P_{\geq 0}
	\end{alignedat}\label{eq:mps-scheme-packing-lp}
\end{equation}
Conveniently, \eqref{eq:mps-scheme-packing-lp} is a \emph{packing LP}: All variables and coefficients are nonnegative, and all constraints are upper bounds.
Such linear programs can be approximately solved very efficiently:
\begin{theorem}[\textcite{packing-lp-solver}] \label{thm:packing-lp-solver}
	There exists a randomized algorithm which, given a packing LP and $\varepsilon>0$,
	computes with constant probability a $(1+\varepsilon)$-approximate solution in
	time $\tilde{\mathcal{O}} \left( \frac{N}{\varepsilon} \right)$, where $N$ is the number of nonzero entries of the constraint matrix. \qed
\end{theorem}
Since we modify the LP \eqref{eq:mps-scheme-lp} to \eqref{eq:mps-scheme-packing-lp} and only compute
approximate solutions, we run a simple post-optimization routine after the MPS scheme,
where we modify at most two entries of $y$ at a time, and allow augmenting $\cT$ by exchanging
a single player by another one for tours $T\in\cT$.

\subsection{Practical Results}
\label{subsec:results}

We evaluate our heuristic on randomly generated vehicle routing instances
with Euclidean distances. Each tour may contain at most a certain number of players.
We consider instances with 50 players and a tour capacity of 5 and instances with 1000 players and a tour
capacity of 50. The heuristic has also been applied successfully to real-world instances in a more complicated
model, including time-dependent travel times, time windows and heterogeneous vehicle fleets.
For all tests, the procedure in \cref{fig:heuristic} was run for 12 iterations with the parameter $\epsilon=0.2$
for \cref{thm:packing-lp-solver}, and 8 parallel threads. 
The post-optimization described at the end of the previous section is only active during the last 6 iterations.

\begin{figure}[ht]
	\centering
	\begin{subfigure}[t]{0.47\textwidth}
		\centering
		\includegraphics[width=\textwidth]{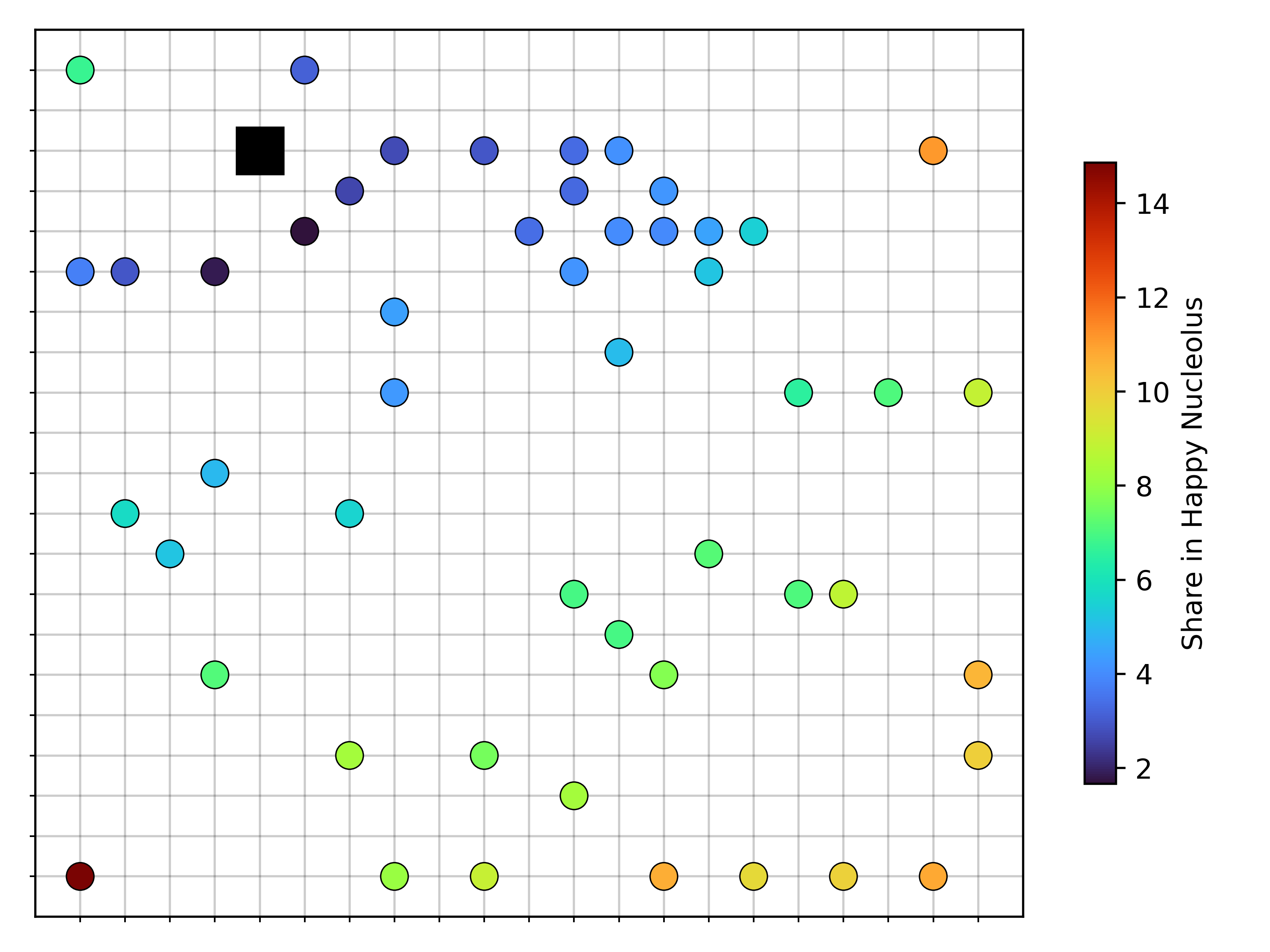}
		\caption{
			The exact happy nucleolus of the instance.
		}
		\label{fig:example-veh-routing-happy-nuc}
	\end{subfigure}
	\hfill
	\begin{subfigure}[t]{0.47\textwidth}
		\centering
		\includegraphics[width=\textwidth]{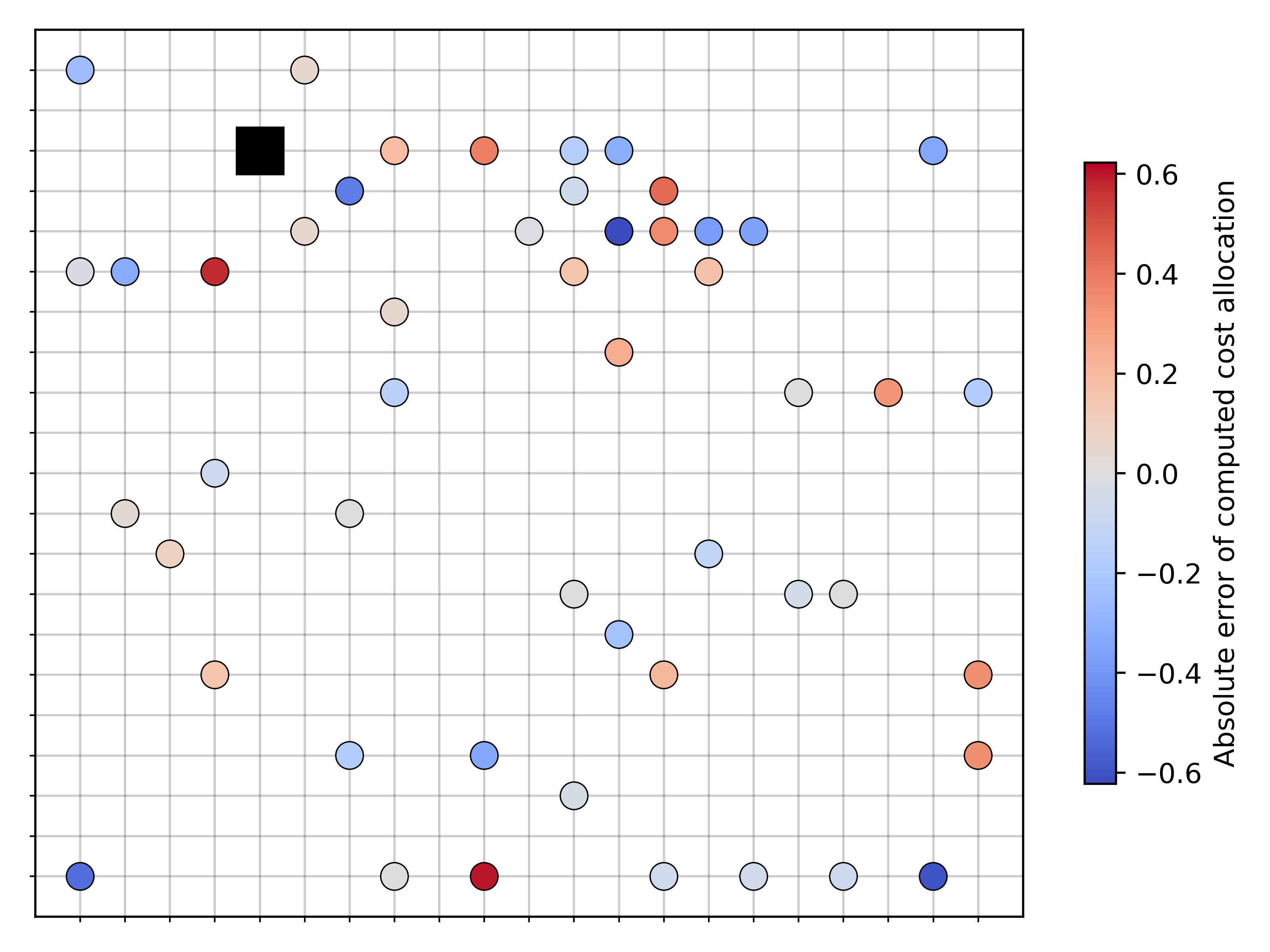}
		\caption{
			The difference between the heuristically computed cost allocation and the
			happy nucleolus.
		}
		\label{fig:example-veh-routing-err}
	\end{subfigure}
	\caption{
		Results for a randomly generated Euclidean vehicle routing instance with depot~(black~square), 50 players (circles) and a tour
		capacity of 5.
	}
	\label{fig:example-veh-routing}
\end{figure}

\begin{figure}[H]
	\centering
	\begin{minipage}{0.47\textwidth}
		\centering
		\includegraphics[width=0.99\textwidth]{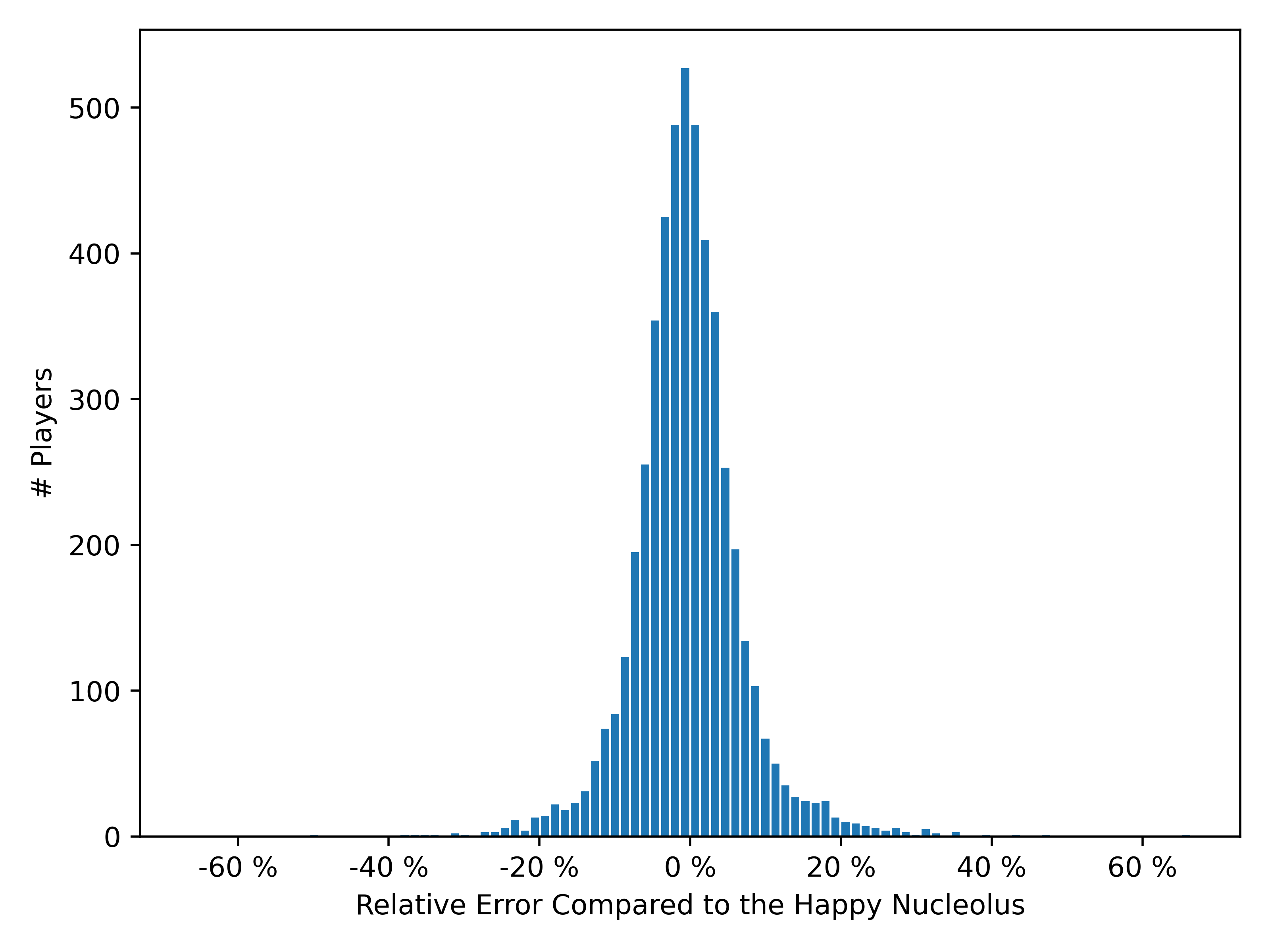}
		\caption{
			Histogram of the relative errors across 100
			vehicle routing instances, each with
			50 players and a tour capacity of~5.\\[-.1cm]
			\phantom{a}
		}
		\label{fig:heuristic-err-distr-hist}
	\end{minipage}%
	\hspace{.5cm}
	\begin{minipage}{0.47\textwidth}
		\centering
		\includegraphics[width=\textwidth]{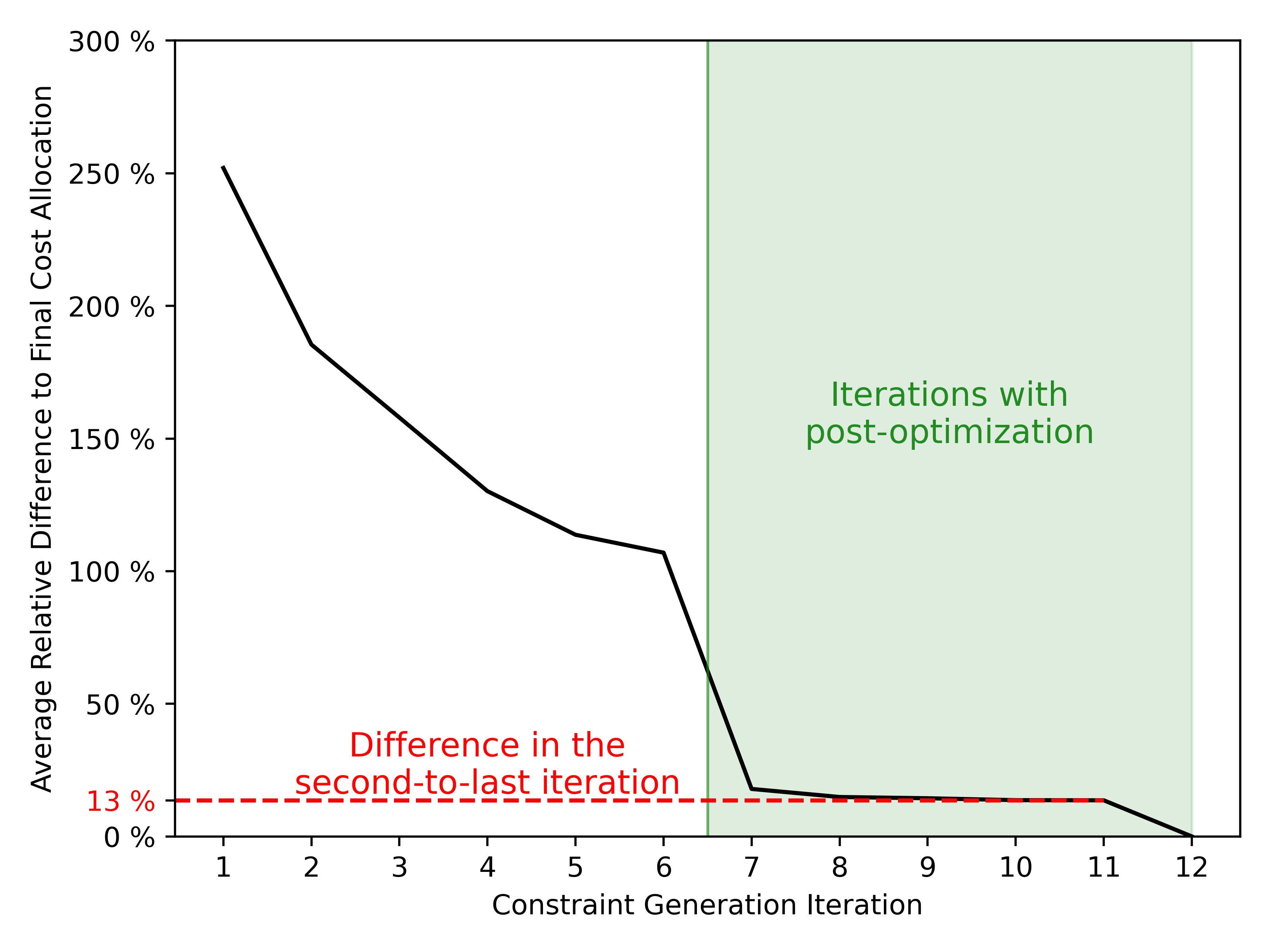}
		\caption{
			Relative difference of the cost allocation in each iteration compared to
			the final iteration for an instance with 1000 players and a tour capacity of 50.
		}
		\label{fig:heuristic-convergence}
	\end{minipage}
\end{figure}

One example is shown in \cref{fig:example-veh-routing}.
\cref{fig:example-veh-routing-err} shows the absolute error of the cost allocation computed by
the heuristic in 2.2 seconds, with an average relative
error of 4.4 \%. The exact happy nucleolus (\cref{fig:example-veh-routing-happy-nuc}) was computed in 413 seconds by enumerating all tours.
The distribution of relative errors across all players of 100 instances of the same size is shown in
\cref{fig:heuristic-err-distr-hist}.

Turning to larger instances with 1000 players, 
our heuristic took roughly 2.5 hours per instance,
while computing the exact happy nucleolus is intractable with current methods. 
Therefore, we cannot compute the error of the heuristic on those instances. However,
\cref{fig:heuristic-convergence} shows that the cost allocation seems to converge to the final
result across the constraint generation iterations and does not change too much in the last few iterations.
While the final result could in theory be far away from the actual happy nucleolus, one would expect the
cost allocation to be less stable in that case. 
We observed this kind of convergence across all our tests.

\section{Future Research Directions} 
\label{sec:future_research}

We have discussed aspects of the relation between nucleolus and happy nucleolus in \cref{sec:relation-nuc-happy-nuc}, 
and disproved general player-wise domination, even for monotone and subadditive games.
As a natural next step, we consider it interesting to study a (multiplicative or additive) bound on the player-wise difference between the two concepts, which cannot exist (instance-independent) for general games, but might exist for monotone and subadditive games.

The discussion in \cref{subsec:mps} might suggest that nucleolus computation is at least as hard as happy nucleolus computation.
This is true when restricting to the MPS scheme:
If we are able to solve \eqref{eq:mps-scheme-lp}, we can compute the happy nucleolus, but not necessarily the nucleolus.
Beyond the MPS scheme, we find a different situation.
For one, there are monotone games for which computing $\HNvalue$ is NP-hard, whereas $\Nvalue$ is trivially given
(e.g., for \emph{weighted voting games} by \cite{bachrach2009cost}).
Even more, there are games for which computing the nucleolus takes polynomial time, but happy nucleolus computation is NP-hard (e.g., for \emph{graph games} by \cite{deng1994complexity}).
It would be interesting to see such an example for a subadditive game, or prove the structural result that nucleolus computation is at least as hard as happy nucleolus computation for the class of monotone and subadditive games.

Further, we are interested in approximability of the happy nucleolus.
For certain classes of games, we have seen how to approximately solve the separation problem of the MPS scheme for happy nucleolus computation in \cref{sec:separation}.
This does not trivially lead to approximation of the happy nucleolus yet, as errors might accumulate.
Is constant-factor approximation even possible? 
For vehicle routing games, the trivial choice of singleton costs does not yield any guarantee.

\section*{Acknowledgements}

We want to thank Jens Vygen for providing the initial draft of \cref{fig:frac_dominated_sets}.

\newpage
\printbibliography

\end{document}